\documentclass{sig-alternate}
\usepackage{url}
\usepackage{multirow}

\newcommand\leadin[1]{%
    \vskip 5pt \noindent\textbf{#1.} %
}

\begin{document}

\date{10 November 2014}

\title{User Session Identification Based on Strong Regularities in Inter-activity Time}

\toappear{This is a non-peer reviewed, pre-submission of an article.}

\numberofauthors{8}

\author{
\alignauthor Aaron Halfaker\\\
       \affaddr{Wikimedia Foundation}\\
       \email{ahalfaker@wikimedia.org}
\alignauthor Os Keyes\\
       \affaddr{Wikimedia Foundation}\\
       \email{okeyes@wikimedia.org}
\alignauthor Daniel Kluver\\
       \affaddr{GroupLens Research}\\
       \affaddr{University of Minnesota}\\
       \email{kluver@cs.umn.edu}
\and  
\alignauthor Jacob Thebault-Spieker\\
       \affaddr{GroupLens Research}\\
       \affaddr{University of Minnesota}\\
       \email{thebault@cs.umn.edu}
\alignauthor Tien Nguyen\\
       \affaddr{GroupLens Research}\\
       \affaddr{University of Minnesota}\\
       \email{nguy1749@umn.edu }
\alignauthor Kenneth Shores\\
       \affaddr{GroupLens Research}\\
       \affaddr{University of Minnesota}\\
       \email{shores@cs.umn.edu}
\and
\alignauthor Anuradha Uduwage\\
       \affaddr{GroupLens Research}\\
       \affaddr{University of Minnesota}\\
       \email{uduwage@cs.umn.edu}
\alignauthor Morten Warncke-Wang\\
       \affaddr{GroupLens Research}\\
       \affaddr{University of Minnesota}\\
       \email{morten@cs.umn.edu}
}

\maketitle
\begin{abstract}
Session identification is a common strategy used to develop metrics for web analytics and behavioral analyses of user-facing systems. Past work has argued that session identification strategies based on an inactivity threshold is inherently arbitrary or advocated that thresholds be set at about 30 minutes. In this work, we demonstrate a strong regularity in the temporal rhythms of user initiated events across several different domains of online activity (incl. video gaming, search, page views and volunteer contributions). We describe a methodology for identifying clusters of user activity and argue that regularity with which these activity clusters appear implies a good rule-of-thumb inactivity threshold of about 1 hour.  We conclude with implications that these temporal rhythms may have for system design based on our observations and theories of goal-directed human activity.

\end{abstract}

\category{H.4}{Information Systems Applications}{Miscellaneous}
\category{H.1.1}{Coding and Information Theory}{Formal models of communication}

\terms{Theory, Measurement, Human Factors}

\keywords{User session, Activity, Human behavior, Regularities, Metrics, Modeling}

\section{Introduction}
\label{sec:introduction}
In 2012, we had an idea for a measurement strategy that would bring insight and understanding to the nature of participation in an online community.  While studying participation in Wikipedia, the open, collaborative encyclopedia, we found ourselves increasingly curious about the amount of time that volunteer contributors invested into the encyclopedia's construction.  Past work measuring editor engagement relied on counting the number of contributions made by a user\footnote{for example, ``Wikipedian is first to hit 1 million edits'' \url{http://www.dailydot.com/news/wikipedian-first-1-million-edits}}, but we felt that the amount of time editors spent editing might serve as a more appropriate measure.

The measurement strategy we came up with was based on the clustering of Wikipedia editors' activities into ``edit sessions'' with the assumption that the duration of an edit session would represent a lower bound of the amount of time invested into Wikipedia contributions~\cite{geiger2013using}.  Through our ethnographic work in Wikipedia we had found the notion of work sessions to be intuitive, yet there did not appear to be a consensus in the literature on how to identify work sessions from timestamped user activities.  This led us to look to the data for insight about what might be a reasonable approach to delineating users' editing activity into sessions.  The regularities we found in inter-activity time amazed us with their intuitiveness and the simplicity of session demarcation they implied. It is that work that led us to look for such regularities in other systems and to write this paper to share our results.

We are not the first to try our hands at identifying a reasonable way to measure user session behavior in human-computer interaction.  User sessions have been used extensively to generate metrics for understanding the performance of information resources~\cite{govseva2006empirical} -- especially in the domain of search~\cite{donato2010you,eickhoff2014lessons} and content personalisation~\cite{lee1999analysis,spiliopoulou2003framework}. Despite this interest in understanding the nature and manifestation of user sessions, no clear consensus about how to perform session identification has emerged.  In fact, some work has gone as far as to argue that sessions don't actually exist as a useful divide for user activity~\cite{jones2008beyond} or that the common strategy of choosing a global inactivity threshold is arbitrary at best~\cite{montgomery2001identifying}.

In this paper, we propose and demonstrate a strategy for identifying user sessions from log data and demonstrate how the results match both intuition and theory about goal-directed human activity.  We also show how this strategy yields consistent results across many different types of systems and user activities.  First, we summarize previous work which attempts to make sense of user session behavior from log data.  Then we discuss theoretical arguments about how goal-directed user behavior ought to manifest in the data.  Third, we discuss a generalized version of the inactivity threshold identification strategy we developed in~\cite{geiger2013using} and present strategies for identifying optimal inactivity thresholds in new data.  Then, we introduce 6 different systems from which we have extracted 10 different types of user actions for analysis and comparison. Finally, we conclude with discussions of the regularities and irregularities between datasets and what that might imply for both our understanding of the measurement of human behavior and the design of user-facing systems

\section{Related work}
\subsection{Human activity sessions}
\label{sec:human_activities}
The concept of an activity session is an intuitive one, but it's surprisingly difficult to tie down a single definition of what a session is, and how it can be demarcated.  A ``session'' may refer to ``(1) a set of queries to satisfy a single information need (2) a series of successive queries, and (3) a short period of contiguous time spent querying and examining results.''~\cite{jones2008beyond}

(1) is referred to, particularly in search-related literature~\cite{eickhoff2014lessons,jones2008beyond}, not as a session but as a task--a particular information need the user is trying to fulfil.  Multiple tasks may happen in a contiguous browsing period, or a single task may be spread out over multiple periods.
(2) is unclear. It may refer to a series of contiguous but unrelated queries (in which case it is identical to the third definition), or a series of contiguous queries based on the previous query in the sequence (in which case it is best understood as a sequence of tasks).
(3) is the most commonly-used definition in the literature we have reviewed~\cite{govseva2006empirical,nadjarbashi2004improving,spiliopoulou2003framework,white2010assessing}. This contrasts with the notion of \emph{task} and is the definition of ``session'' that we have chosen for this paper. It's also the definition used by the W3C~\cite{lavoie1999web}.

We found inspiration in thinking about how to model user session behavior in both the empirical modeling work of cognitive science and the theoretical frameworks of human consciousness as applied to ``work activities''.
 
The lack of purely random distribution in the time between logged human actions has been the topic of recent studies focusing on the cognitive capacity of humans as information processing units.  Notably, Barbasi showed that, by modeling communication activities with decision-based priority queues, he could show evidence for a mechanism to explain the heavy tail in time between activities~\cite{barabasi2005origin} -- a pattern he describes as bursts of rapid activity followed by long periods of inactivity.  Wu et al. built upon this work to argue that short-message communication patterns could be better described by a ``bimodal'' distribution characterized by Poisson-based initiation of tasks and a powerlaw of time inbetween task events\cite{wu2010evidence}.

In contrast, Nardi calls out this cognitive science work for neglecting context in work patterns, motivation and community membership -- thereby inappropriately reducing a human to a processing unit in a vacuum~\cite{nardi1996context} (p21).  Instead, Nardi draws from the framework of Activity Theory (AT) to advocate for an approach to understanding human-computer interaction as a conscious procession of \emph{activities}.  AT describes an activity as a goal-directed or purposeful interaction of a subject with an object through the use of tools. AT further formalizes an \emph{activity} as a collection of \emph{actions} directed towards completing the activity's goal.  Similarly, \emph{actions} are composed of \emph{operations}, a fundamental, indivisible, and unconscious movement that humans make in the service of performing an action.

For an example application of AT, let us examine Wikipedia editing.  Our ethnographic work with Wikipedia editors suggests that it is common to set aside time on a regular basis to spend doing ``wiki-work''.  AT would conceptualize this wiki-work overall as an \emph{activity} and each unit of time spent engaging in the wiki-work as an activity phase -- though we prefer the term ``activity session''.

The \emph{actions} within an activity session would manifest as individual edits to wiki pages representing contributions to encyclopedia articles, posts in discussions and messages sent to other Wikipedia editors.  These edits involve a varied set of \emph{operations}: typing of characters, copy-pasting the details of reference materials, scrolling through a document, reading an argument and eventually, clicking the ``Save'' button.

In this work we draw from both the concepts of the operation-action-activity heirarchy of Activity Theory and the empirical modeling strategies of cognitive science as applied to time between events.

\subsection{Session identification}
\label{sec:session_identification}
User sessions have been used as behavioral measures of human-computer interaction for almost two decades, and for this reason, strategies for session identification from log data have been extensively studied~\cite{eickhoff2014lessons}.

Cooley et al.~\cite{cooley1999data} and Spiliopoulou et al.~\cite{spiliopoulou2003framework} contast two primary strategies for identifying sessions from activity logs: ``navigation-oriented heuristics'' and ``time-oriented heuristics''.

Time-oriented heuristics refer to the assignment of an inactivity threshold between logged activites to serve as a session delimiter.  The assumption implied is that if there is a break between a user's actions that is sufficiently long, it's likely that the user is no longer \emph{active}, the session is assumed to have ended, and a new session is created when the next action is performed. This is the most commonly-used approach to identify sessions, with 30 minutes serving as the most commonly used threshold~\cite{eickhoff2014lessons,spiliopoulou2003framework,ortega2010differences}.  Both threshold and approach appear to originate in a 1995 paper by Catledge \& Pitkow~\cite{catledge1995characterizing} that used client-side tracking to identify browsing behavior. In their work, they reported that the mean time between user observed user events in their data was 9.3 minutes.  They choose to add 1.5 standard deviations to that mean to achieve a 25.5 minutes inactivity threshold.  Over time this proposed inactivity threshold has gradually been smoothed out to 30 minutes.

The utility and universality of this 30-minute inactivity threshold is widely debated; Mehrzadi \& Feitelson~\cite{mehrzadi2012extracting} found that 30 minutes produced artefacts around long sessions, and could find no clear evidence of a global session inactivity threshold\footnote{Note that this conclusion was reached using the same AOL search dataset that we analyze in this paper}, while Jones \& Klinkner~\cite{jones2008beyond} found the 25.5 minute threshold performed ``no better than random'' in the context of intentifying search tasks. Other thresholds have been proposed, but Montgomery and Faloutsos~\cite{montgomery2001identifying} concluded that the actual threshold chosen made little difference to how accurately sessions were identified.

Navigation-oriented heuristics involve inferring browsing patterns based on the HTTP referers and URLs associated with each request by a user. When a user begins navigating (without a referer), they have started a session; when a trail can no longer be traced to a previous request based on the referers and URLs of subsequent requests, the session has ended.  This approach was pioneered by Cooley et al in 2002~\cite{cooley1999data}.  While it demonstrated utility in identifying ``tasks'', and has been extended by Nadjarbashi-Noghani et al.~\cite{nadjarbashi2004improving}, it shows poor performance on sites with framesets due to implicit assumptions about web architecture~\cite{berendt2002impact}. Further, the sheer complexity of this strategy and it's developmental focus on \emph{task} over \emph{session} make it unsuitable as a replacement for time-oriented heuristics in practical web analytics of activity sessions.

In this work, we will challenge the assertion by prior works that (1) no reasonable cutoff is identifiable from the empirical data and (2) a global inactivity threshold is inappropriate as a session identification strategy.  To our knowledge, we are the first to apply a general session identification methodology to a large collection of datasets and conclude that not only are global inactivity thresholds an appropriate strategy for session identification, but also that, for most user-initiated actions, an inactivity threshold of 1 hour is appropriate.

\section{Methods}
\label{sec:methods}
This section is intended to both serve as a description of our methodology as well as to instruct readers on how to apply the same methods to their own datasets.  First, we will discuss how we recommend applying our methodology for identifying inter-activity type component clusters to a dataset.  Then we  describe the origin of our datasets and the cleanup we performed in order to remove artifacts.

\subsection{Fitting inter-activity times}
First, we must gather a dataset of user-initiated actions with timestamps of at least \emph{seconds} resolution.  We generate inter-activity times on a per-user basis, so a relatively robust user identifier is necessary.  While a persistent user identifier such as one associated with a user account is preferable, we have found that in the case of request logs, a fingerprint based on the request's IP and User-agent seems to be sufficient.

Once we have generated per-user inter-activity times, we plot a histogram based on the logarithmically scaled inter-activity time and look for evidence of a valley.  Given the observations we have seen (and report in section~\ref{sec:results_and_discussion}), we expect to see a valley around about 1 hour with peaks around 1 minute and 1 day.  It is at this time that anomalies in the data should be detected and removed.  For example, we found that the time between Wikimedia Mobile Views (described in the next section) had an absurd spike at exactly 18 minutes of inter-activity time caused by a few (likely automated) users and removed their activities from the dataset.

Next, we try to fit a two component gaussian mixture model using expectation maximization~\cite{benaglia2009mixtools} and visually inspect the results\footnote{Note that we tried several strategies for statistically confirming the most appropriate fit -- of which we found Davies--Bouldin index(DBI)~\cite{davies1979cluster} to be most reasonable -- but none were as good as a simple visual inspection, so we employ and recommend the same.}  When the simple bimodal components did not appear to fit the data appropriately, we explored the addition of components to the mixture model with careful skepticism and repeated visual inspection.

Finally, if we have found what appears to be an appropriate fit, we identify a theoretically optimal inter-activity threshold for identifying sessions by finding the point where inter-activity time is equally likely to be within the gaussians fit with sub-hour means (within-session) and gaussians fit with means beyond an hour (between-session).

\subsection{Datasets}
To test this approach to session identification, we used a variety of datasets covering multiple sites, user groups and types of action.

\leadin{Wikimedia sites.} One of the broadest groups of datasets comes from the Wikimedia websites (such as Wikipedia) and covers both page views (read actions) and edits. For the page views, we gather three datasets, each consisting of randomly-sampled page view events from the Wikimedia request logs. These covered app views (page views from the Wikimedia's official mobile app), mobile views (page views to the mobile site) and desktop views (page views to the desktop site). 100,000 IP addresses (or UUIDs, in the case of the app, since it has those built in) were selected, and all requests from those IPs/UUIDs for the month of October 2014 were retrieved. For desktop and mobile views, a UUID was produced by hashing the IP address, the User agent, and the accept\_language provided with each request. After filtering out known crawlers and automata using tobie's ua-parser\footnote{\url{https://github.com/tobie/ua-parser}}, we arrived at three page view datasets consisting of 2,376,891, 932,754 and 2,285,521 pageviews, respectively. These came from 100,000, 235,067 or 247,269 UUIDs. We also extracted inter-edit times from the English Wikipedia using the methodology we employed in~\cite{geiger2013using} -- randomly selecting 1 million edits from 157,342 registered users.

\begin{table*}
\centering
\caption{Fit and threshold information for clusters.  Note that fits correspond to logarithmically scaled (base 2) seconds between events.  For example, $2^{6.7} = 104 \text{ seconds}$.  It's important to report these values in log scale because, while the mean can be re-exponentiated, the standard deviation of log values doesn't make sense that way. }
\begin{tabular}{|r|r|r|r|r|r|r|r|r|r|r|r|r|r|} \hline
\multirow{2}{*}{\textbf{dataset}} &
\multirow{2}{*}{\textbf{theshold (min)}} &
\multicolumn{3}{|c|}{\textbf{short within}} &
\multicolumn{3}{|c|}{\textbf{within}} &
\multicolumn{3}{|c|}{\textbf{between}} &
\multicolumn{3}{|c|}{\textbf{break}} \\ \cline{3-14}
& &
                           $\mu$       & $\sigma$   & $\lambda$
                         & $\mu$       & $\sigma$   & $\lambda$
                         & $\mu$       & $\sigma$   & $\lambda$
                         & $\mu$       & $\sigma$   & $\lambda$\\ \hline
aol search    & 115      &&&
                         & 6.7         & 2.9         & 0.70
                         & 16.8        & 2.2         & 0.30&&&\\ \hline
cyclo. route  & 89       &&&
                         & 5.0         & 2.5         & 0.87
                         & 18.6        & 3.1         & 0.13&&&\\ \hline
wiki. app     & 29       &&&
                         & 5.2         & 2.3         & 0.74
                         & 15.7        & 2.5         & 0.26&&&\\ \hline
wiki. mobile  & 50       &&&
                         & 6.4         & 2.6         & 0.65
                         & 15.8        & 2.5         & 0.35&&&\\ \hline
wiki. desktop & 46       &&&
                         & 5.5         & 2.6         & 0.75
                         & 15.7        & 2.5         & 0.25&&&\\ \hline
osm change    & 101      &&&
                         & 8.6         & 2.1         & 0.68
                         & 15.5        & 2.5         & 0.30
                         & 22.7        & 2.0         & 0.02\\ \hline
wiki. edit    & 80       &&&
                         & 6.8         & 2.5         & 0.83
                         & 15.4        & 2.7         & 0.16
                         & 22.6        & 1.9         & 0.01\\ \hline
mov. rating   & 33       & 3.0         & 1.3         & 0.58
                         & 5.2         & 1.9         & 0.34
                         & 18.0        & 3.0         & 0.07&&&\\ \hline
mov. search   & 52       & 4.0         & 0.8         & 0.30
                         & 5.7         & 2.5         & 0.50
                         & 17.1        & 3.1         & 0.20&&&\\ \hline
lol game      & 14       &&&
                         & 8.3         & 0.5         & 0.59
                         & 14.1        & 2.8         & 0.41&&&\\ \hline
s. o. answer  & 91       &&&
                         & 10.2        & 1.7         & 0.30
                         & 16.6        & 2.9         & 0.63
                         & 23.0        & 1.5         & 0.06\\ \hline
s. o. quest.  & 335      &&&
                         & 12.7        & 1.7         & 0.10
                         & 18.5        & 2.1         & 0.63
                         & 22.4        & 1.7         & 0.26\\ \hline

\end{tabular}
\label{tab:fits}
\end{table*}

\leadin{AOL search} Contrasting with the Wikimedia datasets we used the (now infamous) AOL search logs\footnote{These logs are controversial due to their inclusion of search terms containing private information, and there has historically been an ethical debate about their use. We are confident, however, that our usage does not have ethical implications; we modified the dataset to strip search terms so that it consists solely of unique IDs and timestamps, as has been used in the past.\cite{mehrzadi2012extracting}  See \url{https://en.wikipedia.org/wiki/AOL_search_data_leak} for more discussion.} (aol, search) consisting of 36,389,567 search actions from 657,427 unique IDs. These actions span from March through May of 2006.

\leadin{Cyclopath} We also gathered a dataset from Cyclopath, a computational geowiki leveraging cyclist knowledge~\cite{priedhorsky2008computational}.  The dataset consists of HTTP requests to the Cyclopath server that are automatically labelled by type.  We filtered these requests to include only those that represent a request for a cycle route between two points (cyclopath, route get). This came to 6,123 requests from 2,233 distinct registered users.

\leadin{Movielens} To explore different types of search and contributory behavor, we also extracted logs from the MovieLens movie recommender system, which has been in use since 1997. As of November 2014 there are 225,543 unique users who have provided more than 21 million movie ratings for more than 25,000 movies. From MovieLens, we extracted two datasets: (movielens, rating) consists of movie rating actions from between 1997 until 5 November 2014, and (movielens, search) consists of search actions from 19 December 2007 to 1 January 2014.

\leadin{StackOverflow}. This popular question/answer system relating to programming and software engineering regularly releases public data dumps. For our analysis, we extracted questions asked and answers posted between July 2008 and September 2013. The question dataset (stack overflow, question) consists of 6,397,301 questions from 1,191,748 distinct users, while the answer dataset (stack overflow, answer) consists of 11,463,991 answers from 790,713 distinct users.

\leadin{OpenStreetMap (OSM)} This open-source alternative mapping service also publishes regular database dumps. We downloaded a full history dump of OSM contributions as of 24 February 2014, restricting this to the North American region as defined by Geofabrik\footnote{\url{http://download.geofabrik.de/north-america.html}}, which consists of the United States, Canada and Greenland. OSM groups individual changes to the map into \textit{changesets}\footnote{\url{http://wiki.openstreetmap.org/wiki/API_v0.6#Changesets_2}} when an editor saves their work. We used the timestamp of the last revision in a changeset as the time that the user saved the changeset. The resulting dataset (osm, changeset) contains 13,388,923 million changesets from 46,595 distinct users.  We found that more than 75\% of changesets occured less than 5 seconds of inter-activity time.  We assumed that these represent a data import that set changeset timestamps to the same value and filtered them from the dataset.

\leadin{League of Legends} This widely-played online multiplayer game supports an extension that adds a rating system for users and logs games and play times for the wide set of users of the extension.  Notably, we used this dataset in previous work to study the effect of deviant behaviour on player retention~\cite{shores2014identification}. We took this dataset - consisting of roughly 2.5 million unique players participating in almost 166 million games - and extracted the time between when a user finished a game and started playing the next game (lol, game). Though not all games were captured (see~\cite{shores2014identification} for more details), missing data is believed to be most prevalent around newer players with less consistent play habits.

Taken together, these datasets represent seven different systems and include different interaction mechanisms (mobile apps, mobile devices, desktop devices and a video game interface), and different classes of interaction (web search \& route finding, contributions to collaboratively edited artifacts, page reads, and games played).

\section{Results \& discussion}
\label{sec:results_and_discussion}
In this section, we present and discuss the result of applying our proposed inactivity threshold identification analysis to the datasets.  We start with the common, bimodal cluster fits.  Then we move to more complicated fits and discuss the implications of additional clusters.  Finally, we demonstrate datasets with less suitable fits and discuss what this implies about the nature of participation in these systems.  Reference table~\ref{tab:fits} for fitted values and thresholds.

\subsection{Simple bimodal fits}
\begin{figure}
\centering
\includegraphics[width=.45\textwidth]{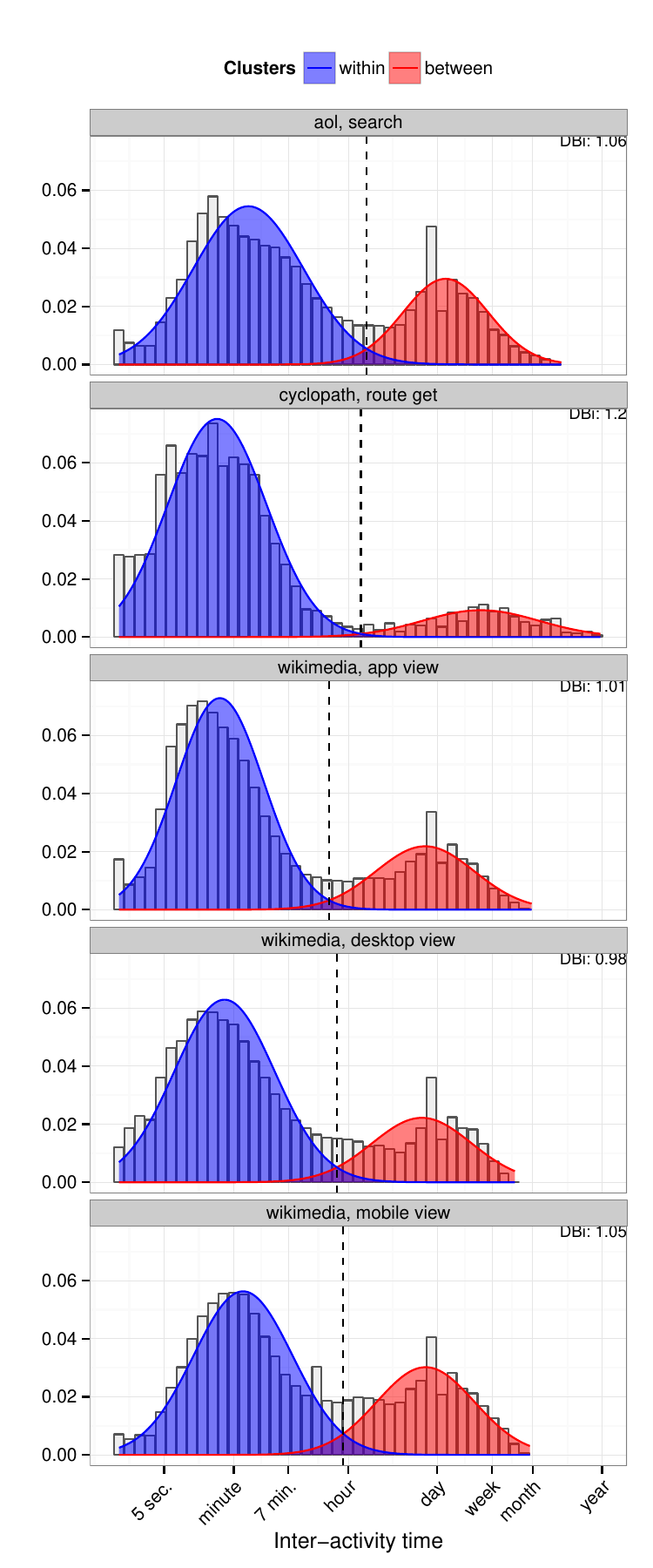}
\caption{
    \textbf{Bimodal clusters.} Empirical inter-activity density (bars) and fitted mixture models of gaussians are plotted for datasets where two clusters appeared to sufficiently explain the observed data.
}
\label{fig:bimodal_clusters}
\end{figure}
Most of the datasets of user-initiated inter-activity times that we observed display a simple bimodal distribution when their histograms are plotted on a logarithmically scaled X axis.  Figure~\ref{fig:bimodal_clusters} plots a log inter-activity time histogram overlaid with expectation maximization fits of a mixture of two log-normal cluster components.  Notably, the AOL search logs represent one of the most clear fits to this bimodal distribution.  This suggests that, counter to Mehrzadi \& Feitelson's conclusions~\cite{mehrzadi2012extracting}, there does seem to to be a clear location for an inactivity cutoff in this dataset -- at approximately one hour.

Figure~\ref{fig:bimodal_clusters} also demonstrates the striking regularity of inter-activity time clusters between systems.  All of the systems presented show a clear fit for a theoretical \emph{within-session} cluster with a mode around one minute and a theoretical \emph{between-session} cluster with a mode at one day.  Each fit intersects at approximately one hour -- with Wikimedia app views displaying the lowest intersection at 29 minutes while AOL searches display the highest intersect at 115 minutes -- nearly two hours.   Despite this variance in the intersection points, a visual inspection of the empirical distribution does not suggest that the choice of a one hour cutoff for either of these datasets would be inappropriate.  Indeed, many of the \emph{between-session} clusters appear to be left shifted due to a lack of longitudinal data, and it is only in these cases that the intersection falls below the one hour mark.

Also of note in these results is the spike of probability of a 24 hour inter-activity time for all but the cyclopath dataset.  This suggests that, for reading Wikimedia sites and searching in AOL, there is a strong tendency to return on a daily basis.  The curious lack of such a day-spike for cyclopath route searches could be explained by the type of usage the site sees. Bicycle route searching may be less of a daily information need than web search and Wikimedia's encyclopedia content.

\subsection{Fits with extended breaks}

\begin{figure}
\centering
\includegraphics[width=.45\textwidth]{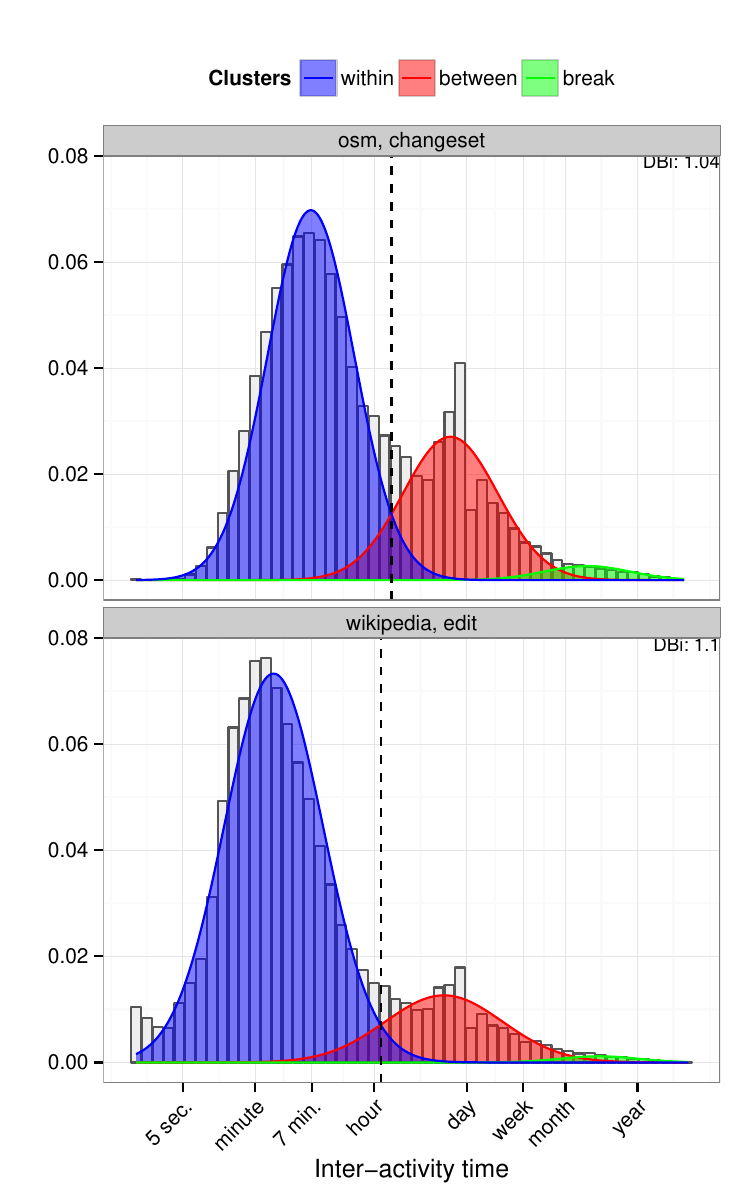}
\caption{
    \textbf{Trimodal clusters.} Empirical inter-activity density (bars) and fitted mixture models of gaussians are plotted for datasets where an additional, ``break'' cluster was needed to fit the data.
}
\label{fig:trimodal_clusters}
\end{figure}

In some cases, we found that the data were fit better by adding a third component to the mixture model that represents very low frequency events.  Figure~\ref{fig:trimodal_clusters} shows the fits for the inter-activity time between OpenStreetMap's changesets and English Wikipedia edits.  Note that, like the bimodal fits above, we again see modes for the \emph{within-session} cluster around one minute and modes for the \emph{between-session} cluster around one day.  However, we found that we could more cleanly fit these datasets with an additional cluster with a mode of around 2.5 months.

As we noted in~\cite{geiger2013using}, we believe that this low frequency cluster represents an extended break from contributing that corresponds to a life event -- like getting married, buying a house, going to school or getting a job.  Wikipedia editors refer to this phenomena in volunteer participation as a ``wikibreak''\footnote{\url{https://en.wikipedia.org/wiki/Wikipedia:Wikibreak}}.  We suspect that the reason for the tiny scale of this cluster is two-fold: (1) contributors who work on Wikipedia or OpenStreetMap for long enough to take an extended break are rare compared to the rate of higher frequency activity and (2) breaks at the scale of 2-3 months often result in total abandonment of participation in the project.

\subsection{Fits with a high frequency component}
\begin{figure}
\centering
\includegraphics[width=.45\textwidth]{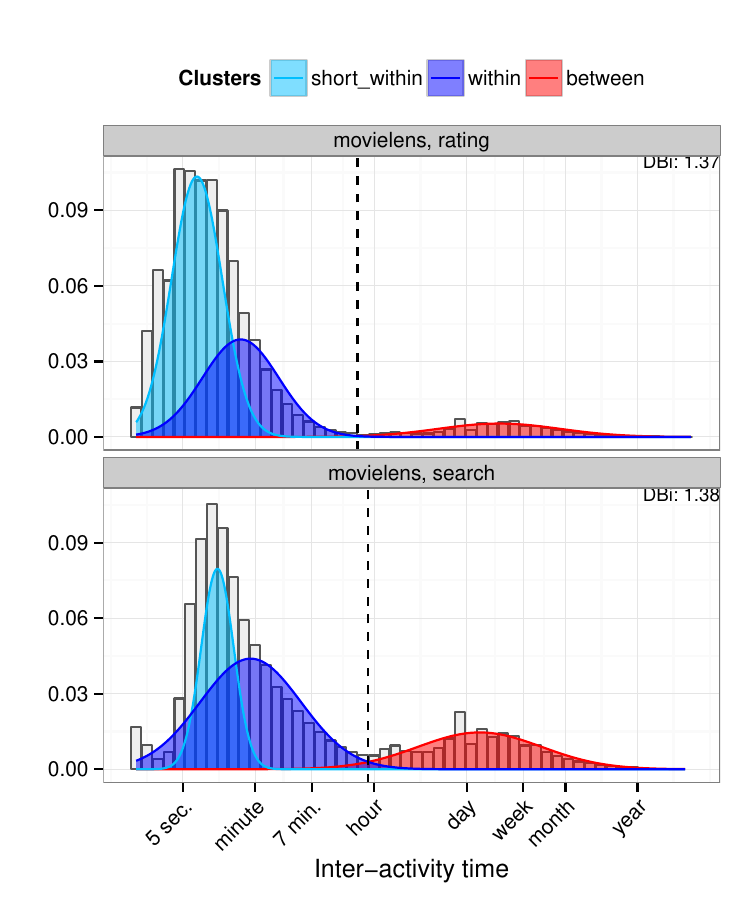}
\caption{
    \textbf{High frequency activity clusters.} Empirical inter-activity density (bars) and fitted mixture models of gaussians are plotted for datasets where an additional, high-frequency inter-activity cluster was needed to fit the data.
}
\label{fig:operation_mixed_clusters}
\end{figure}

When observing the distribution of inter-activity times for ratings and searches in Movielens, we found that both these events occurred with higher frequency than the other datasets.  This made us suspect that there could be an additional cluster component at a high frequency time interval.  Figure~\ref{fig:operation_mixed_clusters} shows how the two datasets lent themselves to this additional ``short within'' component.  Like in previous mixture models, we see a within-session cluster with a mode around one minute and a between-session cluster with a mode around one day.  However, in these datasets we also observed a pattern in inter-activity times that suggested a faster component with a mode around 15 seconds.

Given that this component occurs at shorter intervals than the within-session component, we assume that it also represents within-session activity.  In the case of rating, this high frequency component could represent the rapid rating behavior that the MovieLens interface affords -- a user can rate several movies from a list without leaving a page.  However, we are less sure on how to explain the high frequency component of MovieLens searches.  It could be that, unlike when performing a web search (AOL) or reading encyclopic content (Wikimedia), users' movie searches are more likely to benefit from more rapid iteration.

\subsection{Unusual fits}
\begin{figure}
\centering
\includegraphics[width=.45\textwidth]{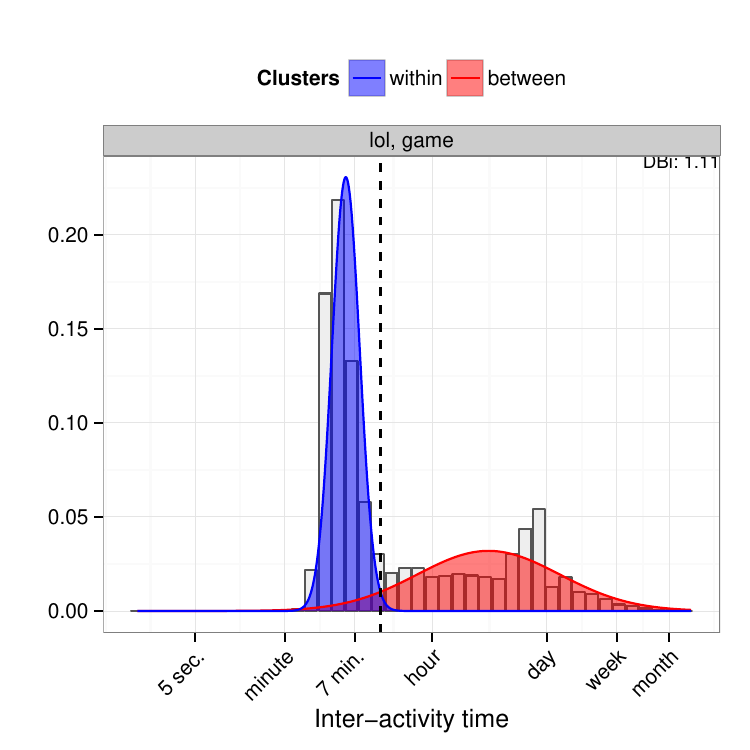}
\caption{
    \textbf{Inter-game clusters.} Empirical inter-activity density (bars) and fitted mixture models of gaussians are plotted for time between League of Legends games.
}
\label{fig:lol_game_clusters}
\end{figure}
While the fits described so far follow a clear pattern with somewhat minor nuance as to the nature of the gaussian fitting strategy, the other datasets we observed suggest that the this strategy for identifying session thresholds is not universally suitable for all user-intiated events.

\leadin{League of Legends}
Figure~\ref{fig:lol_game_clusters} shows the two cluster fit for League of Legends game playing.  Here, we see a very high density component with a mode around five minutes and a very wide component with a mode around five hours.  The intersection of these components place the threshold at approximately 14 minutes.  It is important to note that the tightness of the dense component may be an artifact of the way that inter-game times differ from the inter-activity times observed in the other datasets.  In the case of this dataset, only the time between games is accounted for -- the time between the end of one game and the beginning of the next.

There also may be constraints inherent to the system that limit the potential time spans in which a user could possibly act.  For example, League of Legends employs a queuing mechanism for matching teammates with opponents that takes approximately 5 minutes to complete most of the time.  Our own experience with the game suggests that many users will often finish one game and immediately get into the queue for another.  It is likely that these system limitations are the reason for infrequent between-game times under 1 minute.  It seems clear from this result that understanding a system's limitations on user behavior is important when interpreting cluster fit.

\leadin{Stack overflow}
\begin{figure}
\centering
\includegraphics[width=.45\textwidth]{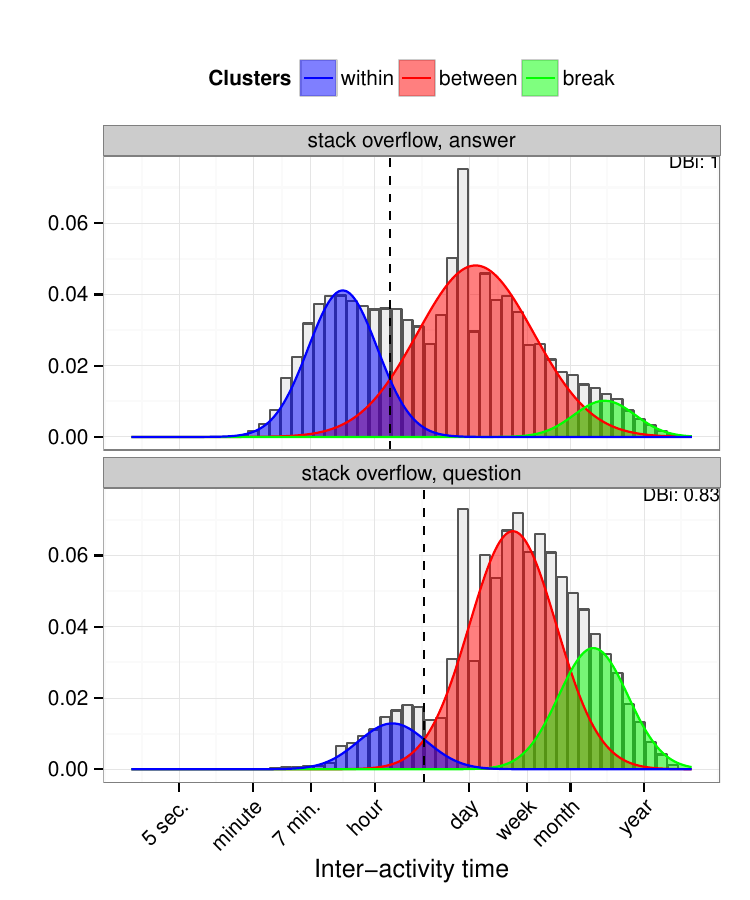}
\caption{
    \textbf{Low frequency clusters.} Empirical inter-activity density (bars) and (non-convergent) fitted mixture models of gaussians are plotted for time between posts on Stack Overflow.
}
\label{fig:stack_overflow_clusters}
\end{figure}
Unlike the other datasets observed, the time between Stack Overflow posts does not suggest a clear valley from which to draw intuition about where to draw a session cutoff.  Figure \ref{fig:stack_overflow_clusters} shows the (non-convergent) fits of question asking and answering activities.  In this case, there is a dramatic reduction in the scale of the higher frequency time components and what appears to be a shift of the within-session component to the right.

If we are to interpret the fit of these clusters as meaningful, the right shift of the within-session component could be due to the time needed to produce a high quality question or answer.  Stack Overflow's incentive structure is designed to encourage high quality posts.  High quality posts are more likely to be reviewed positively by other users, and a user's score within Stack Overflow is largely dependent on how other users rate the quality of their posts\footnote{\url{http://meta.stackexchange.com/help/whats-reputation}}.  It seems likely that producing a high quality post would take a substantial amount of time and that this time investment would make it difficult to complete posts with a high enough frequency to produce a short inter-activity time component like we saw in other systems.  In this case, it seems that either our strategy for identifying a suitable inactivity threshold is insufficient or that Stack Overflow users rarely post more than one question or answer within an activity session.

\vskip 20pt

\section{Implications \& future work}
\label{sec:implications_and_future_work}
In this paper, we have challenged previous literature that suggests no apparent global inactivity threshold exists for identifying user sessions from logs.  From our results, we propose a simple, yet apparently robust, rule of thumb and a methodology for checking this rule in other datasets.  The rule of thumb is easy to apply; our analysis suggests that setting an inactivity threshold to demarcate the end of a session at \emph{one hour} will be appropropriate for most kinds of activity log analysis.

We suspect that this strategy will be robust to new datasets since it is (1) grounded in empirical observations of a natural valley in activity times that corresponds to our intuitions about user activities and (2) holds constant across a wide range of systems and activity types. Even when our threshold detection strategy deviated from one hour, the devations were relatively small given the scale of activities, and in some cases, this deviation could be explained by limitations in the data used to fit our models.  However, we still advise that any new application of session identification using an hour as an inactivity threshold is preceeded by a plot of a histograph of log-scaled inter-activity times and visual inspection for a natural valley between 1 minute and 1 day.

These results and our recommendations stand in the face of a long and nuanced discussion of the nature of user sessions as can be extracted from logged interactions with a computer system.  We place our criticisms of previous work into two categories: (1) previous empirical work did not attempt to look for log-normally distributed patterns and therefore concluded that no obvious separation between within- and between-session inter-activity times exist\cite{mehrzadi2012extracting}\cite{catledge1995characterizing} and (2) other work exploring \emph{task driven} behavior conflates ``task'' with ``session''.  We challenge (1) on the basis of the clear trends represented in the results of this work and (2) by drawing a distinction between goal-directed tasks and activity sessions which often represent a collection of heterogenious goal-directed tasks.

Further, given the strong regularities we see between different types of human-computer interactions, our results suggest something more fundimental about human activity itself.  As discussed in section \ref{sec:human_activities}, Activity Theory(AT) conceptualizes human consciousness as a sequence of \emph{activities} which represent a heirarchical relationship with \emph{actions} and \emph{operations}.  We suspect that the fact that \emph{operations} and \emph{actions} must be performed in a sequence suggests the temporal rhythm we observe.  While it's hard to say conclusively, we suspect that the ``short\_within'' clusters we observe represent \emph{operation}-level events, the ``within'' clusters represent \emph{action}-level events, and the ``between'' clusters represent \emph{activity}-level events.

If this application of AT to the observed patterns is accurate, this could have substantial implications for the design of systems.  System designers may be able to take advantage of the regularities observed by constructing systems that afford operations, actions and activity sessions at timescales that humans will feel find natural.  Our analysis suggests that operations should exist at the timescale of about 5-20 seconds, actions should be completable at a timescale of 1-7 minutes and activities should be supported at daily to weekly time intervals.  We suspect that systems that do not allow users to work at these time scales may be frustrating or may otherwise limit the ability of their users to function at full capacity.

These ruminations about human behavior and its manifestation in well designed systems are only speculation at this point.  New work will need to be done to explore whether our predictions hold and whether limiting or enabling certain types of activity rhythms substantially affects user experience or performance.

\section{Acknowledgments}
\label{sec:acknowledgements}
We thank Stuart Geiger for his involvement in our previous work and for the inspiration he provided toward pushing for measurements that more accurately represent human activity.  This work has been funded in part by the National Science Foundation (grants IIS-0808692, and IIS-1111201).  We are grateful to Dror Feitelson for agreeing to share the AOL search dataset with us. Individual authors would also like to thank Katie Horn and Margret Wander for their feedback and inspiration.

\bibliographystyle{abbrv}
\bibliography{refs}

\end{document}